\documentclass[%
reprint,
superscriptaddress,
amsmath,amssymb,
]{revtex4-1}
\usepackage{color}
\usepackage{graphicx}
\usepackage{dcolumn}
\usepackage{bm}
\usepackage{textcomp}

\usepackage{braket}

\usepackage{comment}
\usepackage{here}
\usepackage[inline]{enumitem}

\begin {document}

\title{Observation-Time-Induced Crossover from Fluctuating Diffusivity}

\author{Masahiro Shirataki}
\affiliation{%
  Department of Physics and Astronomy, Tokyo University of Science, Noda, Chiba 278-8510, Japan
}%

\author{Takuma Akimoto}
\email{takuma@rs.tus.ac.jp}
\affiliation{%
  Department of Physics and Astronomy, Tokyo University of Science, Noda, Chiba 278-8510, Japan
}%



\date{\today}

\begin{abstract}
 A sharp change in apparent mobility at a characteristic temperature that depends on the observation time has been reported in experiments and simulations of hydrated proteins. 
 Such behavior is often discussed in the context of the protein dynamical transition, yet its general physical origin remains unclear.
Here we show that fluctuating diffusivity within a Langevin framework naturally gives rise to an observation-time-induced crossover in translational diffusion: the effective diffusion coefficient exhibits a temperature-dependent change whose crossover point systematically shifts with the observation time. Through analytical and numerical analyses, we elucidate the mechanism of this crossover and identify the minimal conditions required for its emergence.
Our results establish observation-time-induced crossover as a generic non-equilibrium phenomenon in systems with slowly relaxing mobility fluctuations. While distinct from internal dynamical transitions probed in neutron scattering, this framework provides a unified perspective that encompasses related finite-time crossover phenomena observed in hydrated proteins and other complex soft-matter systems.
\end{abstract}

\maketitle


\section{Introduction}

Diffusion theory, originating from the work of Einstein and Smoluchowski \cite{Einstein1905,smoluchowski1906kinetischen}, relates particle mobility to thermal fluctuations. According to the Stokes-Einstein relation \cite{Einstein1905,Stokes}, the diffusion coefficient of a spherical particle is given by
\begin{equation}
D = \frac{k_{\rm B} T}{6\pi \eta r},
\end{equation}
where $T$ is the temperature, $\eta$ is the viscosity of the surrounding medium, $r$ is the particle's hydrodynamic radius, and $k_{\rm B}$ is the Boltzmann constant.

In complex and heterogeneous systems, however, the Stokes-Einstein relation should
be regarded as an effective, coarse-grained description.
In concentrated or interacting protein solutions, for example, generalized Stokes-Einstein relations predict that the collective diffusion coefficient depends not only on temperature and viscosity but also on interparticle interactions and concentration, reflecting many-body and thermodynamic effects \cite{Kholodenko1995,yu2005prediction,heinen2012viscosity}.
Such extensions highlight that experimentally extracted diffusion coefficients often represent effective, environment-dependent transport parameters rather than fixed microscopic constants.

In addition to interaction-induced modifications, internal structural fluctuations and environmental heterogeneity can induce
time-dependent variations of the local friction and the effective hydrodynamic
size experienced by a molecule, leading to a mobility that fluctuates on finite
time scales \cite{Akimoto2026}.
Such effects become particularly relevant when transport properties are extracted
from finite observation windows, as is often the case in experiments and simulations
of biomolecular systems.

Hydrated proteins exhibit a sharp increase in the mean-squared displacement (MSD) around 200-240 K, a well-known feature termed the protein {\it dynamical transition} (DT) \cite{doster1989dynamical,Doster1990,Tournier2003,Becker2004,Roh2005, roh2006influence,Becker2003,Vural2013,Doster2008Thedynamical,doster2010protein,Magazu2011,Ngai2017}.  Strikingly, neutron-scattering experiments show that the apparent transition temperature depends on the instrumental resolution  \cite{Magazu2011, Ngai2017, Becker2004, Doster2008Thedynamical}, 
indicating that the DT is not a true thermodynamic transition but a kinetic crossover governed by the experimental time window \cite{doster2005protein}.
The DT is detected through finite-time atomic fluctuations associated with internal
motions, quantified via the elastic incoherent structure factor or an MSD evaluated
over the instrumental time window.
Thus, the apparent transition temperature reflects the finite-time nature of the measurement and has been widely interpreted as being associated with the onset of additional internal motions becoming accessible within the experimental time window.

Fluctuating diffusivity arises naturally in protein diffusion, where complex conformational dynamics---including long-term memory and folding-unfolding transitions---continuously modulate the hydrodynamic radius \cite{yang2003protein,Kou2004,Min2005,Yamamoto2014b,hu2016dynamics}. 
A Stokes-Einstein-like relation \cite{Yamamoto2021}  implies that these structural fluctuations directly generate temporal variations of the instantaneous diffusivity.
Analogous mechanisms appear in polymer systems: in the reptation model, for example, center-of-mass diffusion is controlled by fluctuations of the end-to-end vector \cite{doi1978dynamics,Doi-Edwards-book,Uneyama2015}, and fluctuating mobility also arises in other coarse-grained polymer models \cite{Miyaguchi2017}.

Such temporal variations of mobility lead to the striking phenomenon of Brownian yet non-Gaussian diffusion (BYNGD), in which the MSD grows linearly in time but the displacement distribution remains non-Gaussian. BYNGD has been reported in complex environments such as cytoplasm, polymer networks, and glassy systems \cite{wang2009anomalous,wang2012brownian,He2013,Bhattacharya2013,Guan2014,kwon2014dynamics,Miotto2021,Rusciano2022}. The diffusing-diffusivity framework has been extensively investigated in this context, including studies of non-Gaussian displacement distributions, first-passage-time statistics, and ergodicity breaking \cite{Akimoto2026,Chubynsky2014,Chechkin2017,sposini2018random,Miyaguchi2019,Hidalgo-Soria2020,hidalgo2021cusp,Sposini2024PRL,*Sposini2024PRE,Massignan2014,Uneyama2015,Miyaguchi2017,cherstvy2016anomalous,Akimoto2016,AkimotoYamamoto2016a,Miyaguchi2016,wang2020fractional}, establishing fluctuating diffusivity as a central mechanism of anomalous transport.

While most previous studies focused on displacement statistics, we address a different facet of fluctuating diffusivity: the temperature dependence of the effective diffusion coefficient under finite observation time. We show that, under non-equilibrium initial conditions, a sharp crossover in the temperature-dependent effective diffusivity naturally emerges, with the crossover point shifting systematically on the observation time.
This observation-time-induced crossover is distinct from BYNGD and from equilibrium phase transitions. We demonstrate that it arises generically from the interplay of fluctuating diffusivity, temperature-dependent relaxation, and finite-time observation.


To clarify this mechanism, we study a double-well-controlled diffusing diffusivity (DWCDD) model, where the instantaneous diffusivity evolves stochastically in a double-well potential, mimicking conformational transitions between folded and unfolded states. For analytical tractability, we reduce this model to a coarse-grained two-state representation that captures the essential features of the effective diffusivity. Through analytical and numerical approaches, we establish the necessary conditions for the observation-time-induced crossover and provide a minimal framework for interpreting dynamical transitions in complex systems. Our results highlight the importance of observation-time effects in the analysis of transport phenomena in biomolecular and soft-matter experiments.


\section{Model}

\subsection{Langevin equation with fluctuating diffusivity}

In the Langevin equation with fluctuating diffusivity, the diffusion coefficient
$D(t)$ is treated as a stochastic process that varies in time, reflecting temporal
heterogeneity in the medium or slow internal degrees of freedom of the particle.
The particle position $x(t)$ then evolves according to 
\begin{equation}
    \frac{d x(t)}{d t} = \sqrt{2 D(t)}\, \xi(t),
\end{equation}
where $\xi(t)$ is a Gaussian white noise satisfying
\begin{equation}
    \langle \xi(t) \rangle = 0, \quad \langle \xi(t)\xi(t') \rangle = \delta(t - t').
\end{equation}
Although the diffusivity $D(t)$ originates from environmental fluctuations or from internal degrees of freedom of the particle---such as conformational changes---we assume that $D(t)$ and $\xi(t)$ are statistically independent.
This assumption reflects a separation of timescales: slow structured or environmental 
 dynamics modulate the effective mobility and are encoded in $D(t)$, whereas
$\xi(t)$ represents fast, memoryless thermal fluctuations hat act on much shorter timescales.

At a fully microscopic level, weak correlations between mobility
fluctuations and thermal noise may exist, especially when internal conformational
dynamics couple to fast vibrational modes. In the present coarse-grained description,
such correlations are neglected as subleading effects that average out over the
mesoscopic timescales on which $D(t)$ is defined.
The present coarse-grained description is therefore designed to capture the
leading-order transport consequences of slow mobility relaxation, rather than
microscopic noise-mobility correlations.

We emphasize that this description is phenomenological and intended as an effective
finite-time transport model. The fluctuating diffusivity does not represent fluctuations of
the thermodynamic temperature, nor does it replace microscopic equilibrium
statistical mechanics.
Instead, $D(t)$ should be interpreted as a coarse-grained, time-dependent mobility
emerging from unresolved internal or environmental degrees of freedom. The framework
is therefore applicable to systems in which a clear separation exists between fast
thermal fluctuations and slower mobility-modulating dynamics, such as proteins,
polymers, and other soft-matter systems exhibiting slow conformational relaxation \cite{Akimoto2026,Kou2004,wang2009anomalous,Doi-Edwards-book}.

\subsection{Double-well-controlled diffusing diffusivity model}
In the  DWCDD model, the particle's diffusion coefficient $D(t)$ is assumed to follow a Stokes--Einstein-type relation \cite{Yamamoto2021,Kimura2022},
\begin{equation}
D(t) = \frac{k_\mathrm{B} T}{r(t)},\label{eq:SEtype-relation}
\end{equation}
where $T$ is the temperature, $r(t)$ represents the instantaneous gyration radius of the particle, and $k_{\rm B}$ is the Boltzmann constant. 
Here, $T$ denotes the thermodynamic temperature of a single surrounding heat bath.
The fluctuating diffusivity therefore originates from conformational fluctuations of $r(t)$ rather than from temperature variations.

The stochastic dynamics is governed by an overdamped Langevin equation in a double-well potential, modeling conformational fluctuations between compact and extended states \cite{Kimura2022,shimizu2025memory}:
\begin{equation}
\gamma \frac{d r(t)}{d t} = -\frac{\partial  V(r)}{\partial r} + \sqrt{2 \gamma k_{\rm B} T}\, \eta(t),\label{eq:LE_r}
\end{equation}
where $\gamma$ is the friction coefficient, $V(r)$ is a double-well potential, and $\eta(t)$ is Gaussian white noise satisfying
\begin{equation}
\langle \eta(t) \rangle = 0, \quad \langle \eta(t)\eta(t') \rangle = \delta(t - t').
\label{eq:wgn_eta}
\end{equation}
The noise amplitude $\sqrt{2\gamma k_{\rm B}T}$ ensures that the dynamics of $r(t)$ satisfies the fluctuation-dissipation relation at temperature $T$. 
The noise $\eta(t)$ is assumed to be statistically independent of $\xi(t), x(t)$ and $r(t)$.

In this model, we employ a double-well potential of the form
\begin{align}
    V(r) = a r^4 + b r^3 + c r^2 + d r + e.
    \label{eq:DWpotential}
\end{align}
which represents the minimal polynomial form capable of generating two
metastable states separated by an energy barrier.
The parameters are set to $a = 0.05$, $b = -0.6$, $c = 2.3$, $d = -3$, and $e = 1.25$. 
These values are chosen to (i) ensure the existence of two stable minima with
distinct radii $r_+=1$ and $r_-=5$, corresponding to two different
diffusivity states, (ii) control the barrier height and thus the
temperature-dependent relaxation timescale, and (iii) maintain numerical
stability over the temperature range considered.
The potential is symmetric about the barrier at $r_{\mathrm{M}}=3$ for
analytical simplicity; the role of asymmetry is discussed later
in Sec.~III~C.

To avoid unphysical values such as negative or extremely large diffusion coefficients,  
a reflecting boundary is imposed at boundaries at $r = 0.1$ and $r = 7$.  
The lower boundary is physically motivated by interpreting $r(t)$ as the gyration radius of the molecule: when the system reaches a highly compact conformation ($r \sim 0.1$), strong steric repulsion is assumed to arise, preventing further compression and effectively pushing $r(t)$ to expand. The upper boundary at $r = 7$ is introduced to prevent unbounded expansion at high temperatures, ensuring physical plausibility of the conformational fluctuations.

\begin{figure}
        \includegraphics[width=.95\linewidth]{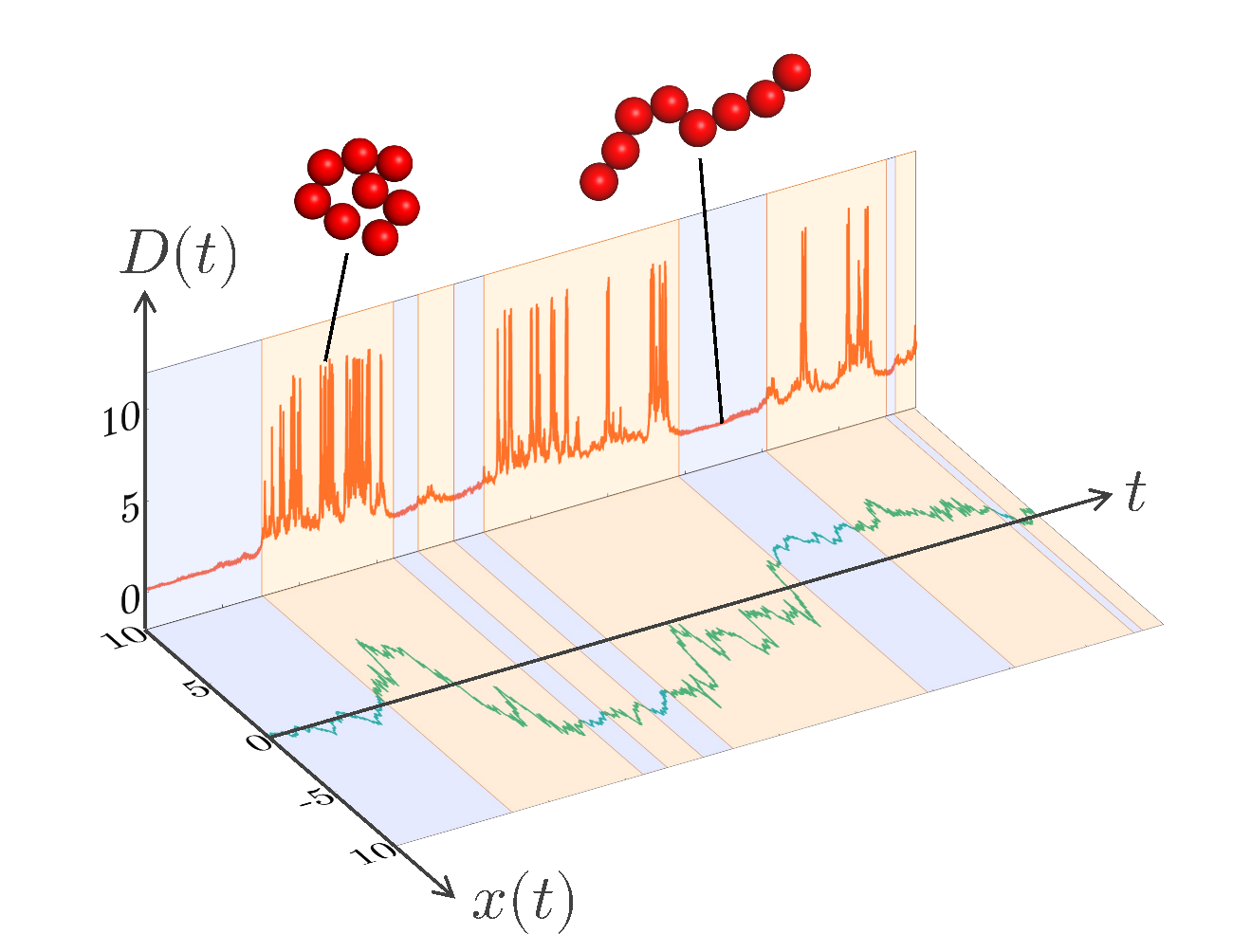}
        \caption{Particle trajectory $x(t)$ and corresponding time-dependent diffusion coefficient $D(t)$ in the  DWCDD model.}
        \label{fig:traj_LEFD}
  \end{figure}

This potential is not intended to quantitatively reproduce a specific
biomolecular free-energy landscape, but to provide a minimal
coarse-grained representation of two metastable mobility states. 
Figure~\ref{fig:traj_LEFD} shows the time evolution of the diffusion coefficient $D(t)$ and the corresponding particle trajectory $x(t)$ in the  DWCDD model.

\subsection{Initial condition}
To explore the DT, we consider non-equilibrium initial conditions in which the system starts in the low-diffusivity state. Specifically in the  DWCDD model, the initial value of the conformational variable is set to  $r(0) = r_\mathrm{-}$, where $r_{\mathrm{-}}$ corresponds to the right minimum of the double-well potential and thus represents the low-diffusivity state. In this way, we can investigate the non-equilibrium character of the MSD and how it reflects the transient dynamical behavior arising from the initial condition.

\subsection{Mapping onto two-state model}

To elucidate its essential features, we describe its dynamics using a coarse-grained two-state representation.
In this description, 
 the diffusivity alternates between two states, denoted as the $+$ and $-$ states, characterized by radii $r_+$ and $r_-$ with $r_- > r_+$. 
The corresponding diffusion coefficients are given by
\begin{equation}
D_\pm = \frac{k_{\rm B} T}{r_\pm},
\end{equation}
which serves as the discrete analogue of Eq.~\eqref{eq:SEtype-relation}.

This two-state picture becomes accurate when the temperature is low compared to the barrier height of the double-well potential ($\Delta U/k_{\rm B}T \gg 1$), so that escape events between the wells are rare and can be treated as statistically independent. 
In this Arrhenius regime, the continuous dynamics of the DWCDD model can therefore be coarse-grained into two discrete diffusivity states. 
Each state persists for a random sojourn time $\tau$, drawn from a state-specific probability density function (PDF) $\rho_\pm(\tau)$.
For analytical tractability, we assume exponential sojourn-time distributions, corresponding to a Markovian switching process. 
While more general distributions may be relevant in complex systems, the exponential case captures the essential features of the observation-time-induced crossover and serves as a natural starting point. Other sojourn-time distributions are discussed in Appendix C.
Specifically, we set
\begin{equation}
    \rho_\pm(\tau) = \frac{1}{\mu_\pm} e^{-\tau / \mu_\pm}, \quad \tau \geq 0,  \label{eq:exp_dist}
\end{equation}
where $\mu_+$ and $\mu_-$ denote the mean sojourn times in the $+$ and $-$ states, respectively.

In the  DWCDD model, the variable $r(t)$ fluctuates within a double-well potential.  
Since the initial condition is set at the right minimum, $r = r_\mathrm{-}$, the system initially remains in the low-diffusivity basin. 
As temperature increases, the particle occasionally acquires enough thermal energy to escape from this well and begins to explore both minima at $r_{\mathrm{-}}$ and $r_{\mathrm{+}}$.
The distribution of $r(t)$ then gradually relaxes toward the equilibrium distribution.
In this regime, the sojourn times in the two wells can be interpreted as temperature-dependent escape times from the corresponding minima.
Thus, the dynamics effectively reduce to stochastic switching between the two discrete states $r = r_{\mathrm{+}}$ and $r = r_{\mathrm{-}}$, which justifies mapping the DWCDD model onto the coarse-grained two-state model.

From the Arrhenius law, the mean sojourn times in the $+$ and $-$  states are given by
\begin{align}
\mu_\pm &\sim 2 \alpha_\pm \delta \pi \exp\left( \frac{V(r_\mathrm{M}) - V(r_\mathrm{\pm})}{k_\mathrm{B} T} \right),
\label{eq:mu_pm}
\end{align}
valid in the activated regime $\Delta U / (k_{\mathrm B} T) \gg 1$, where $\Delta U = V(r_{\mathrm{M}}) - V(r_{\mathrm{\pm}})$ is the barrier height.
Here, $r_{\mathrm{+}}$ and $r_{\mathrm{-}}$ denote the positions of the left and right minima, respectively, and $r_{\mathrm{M}}$ is the position of the barrier top. 
The prefactors $\alpha_\pm$ are determined by the curvatures of the potential near $r = r_{\mathrm{\pm}}$.

\section{Results and Discussion}


\subsection{Observation-time-induced crossover}
In the two-state model, the MSD of a particle, starting from $x(0) = 0$, can be computed as \cite{Miyaguchi2016, Akimoto-Yamamoto2016}:
\begin{align}
     \langle x^2(t) \rangle = 2 \int_0^t \langle D(t') \rangle \, dt' .
 \label{eq:MSD}    
\end{align}
In equilibrium, the ensemble average of $D(t)$ becomes time-independent and equals 
\begin{equation}
D_{\rm eq} = \frac{\mu_+ D_+ + \mu_- D_-}{\mu_+ + \mu_-},
\end{equation}
the stationary mean determined by the equilibrium probabilities of the two states. 

In contrast, under non-equilibrium initial conditions, the ensemble average $\langle D(t) \rangle$ acquires explicit time dependence. It can be expressed as
\begin{align}
    \langle D(t) \rangle = D_- + (D_+ - D_-) p_+(t),\label{eq:Dtwithp+}
\end{align}
where $p_+(t)$ is the probability that the system is in the fast state ($+$) at time $t$.
Since the system is initialized in the slow state, $p_+(0)=0$, and the time evolution of $p_+(t)$ follows from standard renewal-theory results  \cite{Cox, God2001, Miyaguchi2016, Akimoto2023}. 


Starting from the slow state, the time-dependent average diffusivity is given by
\begin{align}
    \langle D(t) \rangle = D_{\rm eq} - (D_+ - D_-)\frac{\tilde{\mu}}{\mu_-} e^{-t/\tilde{\mu}},
\end{align}
where $\tilde{\mu} \equiv \mu_+\mu_-/(\mu_+ + \mu_-)$ is the effective relaxation time of the two-state dynamics. 
Using Eq.~(\ref{eq:MSD}) and defining effective diffusion coefficient as 
\begin{equation}
D_{\mathrm{eff}}(t)\equiv \frac{\langle x^2(t) \rangle}{2t}, 
\end{equation}
we obtain
\begin{align}
    D_{\mathrm{eff}}(t)  = D_{\rm eq} -  (D_+ - D_-)\frac{\tilde{\mu}^2}{\mu_-t} \left(1-e^{-t/\tilde{\mu}}\right).
    \label{eq:D_eff_lambda}
\end{align}
Equation~\eqref{eq:D_eff_lambda} describes how the effective diffusivity evolves from its initial non-equilibrium value toward the equilibrium diffusivity $D_{\rm eq}$.
The first term gives the long-time limit, while the second term quantifies the transient decay originating from the initial slow state.
The crossover is governed by the characteristic relaxation time $\tilde{\mu}(T)$: the effective diffusivity transitions smoothly from its initial to its equilibrium value when the observation time becomes comparable to $\tilde{\mu}(T)$.

By substituting mean escape time $\mu_\pm$ Eq.~\eqref{eq:mu_pm} into Eq.~\eqref{eq:D_eff_lambda}, the effective diffusion coefficient in the double-well-controlled diffusing diffusivity model can be written as
\begin{widetext}
\begin{equation}
    D_\mathrm{eff}(t) = D_- + (D_+ - D_-) \, \frac{\alpha e^{-\frac{V(r_\mathrm{R})}{k_\mathrm{B}T}}}{\alpha e^{-\frac{V(r_\mathrm{R})}{k_\mathrm{B}T}} + \beta e^{\frac{-V(r_\mathrm{L})}{k_\mathrm{B}T}}} \left[1-\frac{1 - \exp\left({-\frac{\exp({-V(r_\mathrm{M})/k_\mathrm{B}T})}{2\alpha\beta\delta\pi}(\alpha e^{-\frac{V(r_\mathrm{R})}{k_\mathrm{B}T}} + \beta e^{\frac{-V(r_\mathrm{L})}{k_\mathrm{B}T}}) t}\right)}{{\frac{\exp({-V(r_\mathrm{M})/k_\mathrm{B}T})}{2\alpha\beta\delta\pi}(\alpha e^{-\frac{V(r_\mathrm{R})}{k_\mathrm{B}T}} + \beta e^{\frac{-V(r_\mathrm{L})}{k_\mathrm{B}T}}) t}} \right].
    \label{eq:D_eff_DWCDDmodel}
\end{equation}
\end{widetext}

Equation~\eqref{eq:D_eff_DWCDDmodel} is not limited to the specific potential given by ~\eqref{eq:DWpotential},  
but can be applied to any double-well potential in the regime where the Arrhenius law holds, i.e., when $k_B T \ll \Delta U$.  

For a symmetric double-well potential centered at $r = r_\mathrm{M}$, substituting the mean escape times $\mu_\pm$ from Eq.~\eqref{eq:mu_pm} into Eq.~\eqref{eq:D_eff_lambda} yields the effective diffusion coefficient of the  DWCDD  model as
\begin{align}
    D_\mathrm{eff}(t)
    = D_- + \frac{D_+ - D_-}{2}
    \left[
        1 - \frac{1 - \exp\!\left(-t/\kappa(T)\right)}
        {t/\kappa(T)}
    \right],
    \label{eq:D_eff}
\end{align}
where $\kappa(T) \equiv \alpha \delta \pi e^{\frac{\Delta U}{k_\mathrm{B}T}}$ is the characteristic relaxation time.
Here, $\Delta U = V(r_\mathrm{M}) - V(r_\pm)$ denotes the barrier height, and symmetry implies
$\alpha \equiv \alpha_+ = \alpha_-$.
Equation~\eqref{eq:D_eff} reveals how the effective diffusivity evolves over time as the system relaxes between the two diffusivity states separated by an energy barrier $\Delta U$.
The temperature dependence of $\kappa(T)$ follows an Arrhenius form, increasing exponentially with $\Delta U$ and decreasing with temperature.
For short observation times $t \ll \kappa(T)$, the expansion $1 - e^{-t/\kappa(T)} \simeq t/\kappa(T)$ gives $D_\mathrm{eff}(t) \simeq D_-$, indicating that the system remains in its initial low-diffusivity state. 
For long times $t \gg \kappa(T)$, the exponential term vanishes and $D_\mathrm{eff}(t)$ approaches the equilibrium value $(D_+ + D_-)/2$.
Thus, Eq.~\eqref{eq:D_eff} captures the observation-time-dependent crossover from non-equilibrium to equilibrium diffusivity, governed by thermally activated transitions between conformational states.

\begin{figure}[t]
  \centering
  \includegraphics[width=\linewidth]{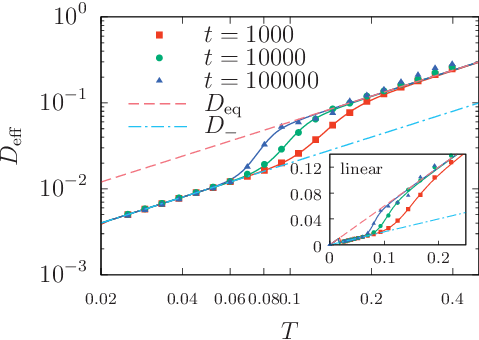}\par\vspace{2mm}
  \includegraphics[width=\linewidth]{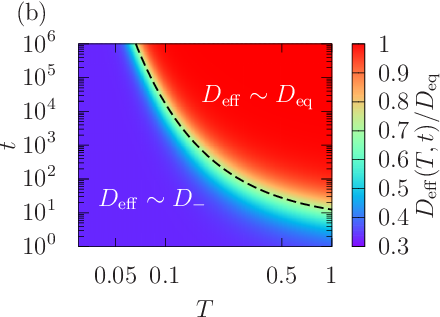}
  \caption{
  (a) Temperature-dependent effective diffusion coefficient for different observation times in the DWCDD model.
  Symbols indicate simulation results, dashed lines represent the initial and equilibrium diffusion coefficients, and solid lines show the theoretical prediction given by Eq.~\eqref{eq:D_eff}.
  Inset: linear-scale zoom around $T<0.2$.
    (b) Heat map of the effective diffusion coefficient $D_{\mathrm{eff}}(T,t)$ in Eq.~\eqref{eq:D_eff}, normalized by the equilibrium diffusion coefficient $D_{\mathrm{eq}}$.
  The black dotted line indicates the escape time of the particle radius $r$ from the potential, given by Eq.~\eqref{eq:mu_pm}.
  }
  \label{fig:DT_and_heatmap}
\end{figure}

Figure~\ref{fig:DT_and_heatmap} (a) shows the temperature dependence of the effective diffusion coefficient $D_{\mathrm{eff}}$ in the  DWCDD model.  
A clear transition in $D_{\mathrm{eff}}$ is observed with increasing temperature, and the transition temperature systematically depends on the observation time.
Theoretical predictions from Eq.~\eqref{eq:D_eff} (solid lines) are in excellent agreement with the simulation results (symbols), confirming that the observation-time-induced crossover arises from thermally activated transitions between the two diffusivity states.

Figure~\ref{fig:DT_and_heatmap} (b) shows a heat map of the effective diffusion coefficient
$D_{\mathrm{eff}}(T,t)$ defined in Eq.~\eqref{eq:D_eff}, normalized by the
equilibrium diffusion coefficient $D_{\mathrm{eq}}$.
The map clearly separates into two regimes, $D_{\mathrm{eff}}\simeq D_{-}$ and
$D_{\mathrm{eff}}\simeq D_{\mathrm{eq}}$, indicating a crossover between the
initial low-diffusivity state and the equilibrated regime.
Importantly, the boundary temperature of this crossover depends on the observation
time, demonstrating the observation-time-induced nature of the transition.
The black dotted line represents the escape time from the double-well potential,
Eq.~\eqref{eq:mu_pm}, showing that the relaxation time of diffusion in this system
is well estimated by the escape time.

While we assume that the dynamics of $r(t)$ are Markovian for analytical simplicity,
it has been demonstrated in experiments and molecular-dynamics simulations that reduced
conformational coordinates of proteins, such as the radius of gyration or the effective
hydrodynamic radius, generally exhibit non-Markovian dynamics \cite{yang2003protein,Kou2004,Min2005,Yamamoto2014b,hu2016dynamics}.
More generally, similar non-Markovian effects arise when spatial heterogeneity,
collective motions, or many-body interactions are coarse-grained into a reduced set of
effective coordinates.
In such cases, the influence of these complex interactions is naturally encoded in
memory kernels and colored noise terms, leading to a generalized Langevin description
of the reduced dynamics.
We have further confirmed that the observation-time-induced crossover persists when the
internal coordinate is modeled by such a generalized Langevin equation, indicating that
the phenomenon is robust against non-Markovian internal dynamics (see Appendix~B).

\subsection{Conditions for the emergence of observation-time-induced crossover}
We now discuss the conditions under which a dynamical transition emerges, based on the results presented above.
Our analysis indicates that the observation-time-induced crossover is governed by the following three conditions:
\begin{enumerate}
\item[(i)] The instantaneous diffusivity fluctuates in time in a stationary manner.
\item[(ii)] The relaxation time of the diffusivity depends on temperature.
\item[(iii)] The system is initially out of equilibrium.
\end{enumerate}
Here, ``stationary" means that the diffusivity dynamics admits a steady state; equilibrium is not required, and non-equilibrium steady states also satisfy (i) (see Appendix~A, where this point is illustrated using a non-equilibrium three-state model).
The second condition is essential, as the apparent transition temperature depends on the observation time; equivalently, the diffusivity must switch between states on a temperature-dependent timescale. Because condition (i) is a prerequisite for the others and is generically satisfied in our framework, our discussion focuses on (ii) and (iii).

\begin{figure*}
  \centering
  \includegraphics[width=\textwidth]{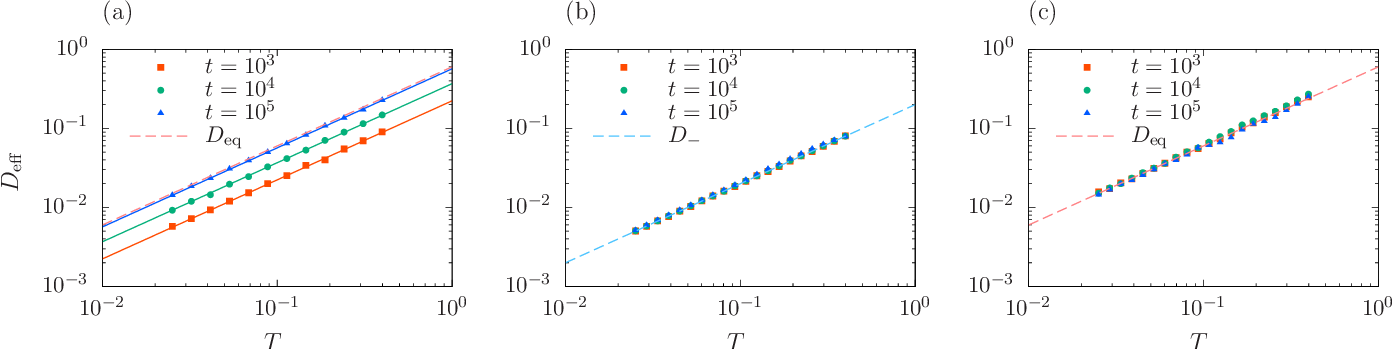}
  \caption{Temperature-dependent effective diffusion coefficient for different observation times. (a) Two-state model, where the mean sojourn time $\mu$ is temperature independent. (b) Diffusing-diffusivity model where the particle radius $r$ evolves under a harmonic potential. (c) Double-well-controlled diffusing-diffusivity model with the initial condition of $r$ sampled from the equilibrium distribution.}
  \label{fig:condition}
\end{figure*}

Figure~\ref{fig:condition} demonstrates the necessity of conditions (i)-(iii) for the emergence of the observation-time-induced crossover, using two representative models.  
Panel (a) shows the result for the two-state model, where the mean sojourn times are set to be independent of temperature.  
In this case, although the instantaneous diffusivity fluctuates and switches between two discrete diffusion coefficients, the switching timescale remains temperature independent.
As a result, condition (ii) is violated, and no observation-time-induced crossover appears.
Panel (b) presents results for the diffusing diffusivity model,  
in which the potential is replaced by a harmonic approximation around $r = r_\mathrm{-}$, eliminating bistability and thus suppressing diffusivity switching.
In this model, transitions between distinct conformational states are absent, and condition (ii) is again not satisfied, and the observation-time-induced crossover does not emerge.
Panel (c) examines the role of non-equilibrium initial conditions, corresponding to condition (iii).  
Here, the system is initialized with $r$ sampled from the equilibrium distribution. In this case, there is no observation-time-induced crossover. 
Because the dynamical transition relies on relaxation from a non-equilibrium state, its emergence requires that the system be initially prepared out of equilibrium. Taken together, these results confirm that both temperature-dependent relaxation and non-equilibrium initial conditions are indispensable for the emergence of the observation-time-induced crossover.

It is worth emphasizing that requirement (iii) distinguishes the present
observation-time-induced crossover from BYNGD.
In many BYNGD settings, non-Gaussian displacement statistics arise even when the
diffusivity process is initialized from its stationary distribution.
By contrast, the crossover studied here is inherently a finite-time,
relaxation-driven phenomenon that becomes pronounced only when the diffusivity
is initially biased away from stationarity.

\subsection{Asymmetric double-well control of diffusing diffusivity}
In this section, we show that by employing an asymmetric potential for the particle radius \(r\) in the  DWCDD model, the theoretical framework of the two-state model can be applied to generic (asymmetric) double-well potentials.
We employ the following composite double-well potential:
\begin{align}
    V(r)=
    \begin{cases}
    U_\mathrm{L}\!\left[\left(\dfrac{r-3}{2}\right)^2 - 1\right]^2, & \text{if } r \le 3, \\
    U_\mathrm{R}\!\left[\left(\dfrac{r-3}{2}\right)^2 - 1\right]^2 + (U_\mathrm{L} - U_\mathrm{R}), & \text{if } r > 3,
    \end{cases}
\end{align}
where \(U_\mathrm{L}\) and \(U_\mathrm{R}\) control the well depths on the left and right, respectively. This potential has minima at \(r_+=1\) and \(r_-=5\), and a potential barrier at \(r_\mathrm{M}=3\).
By replacing the potential in Eq.~\eqref{eq:DWpotential} of the main text with the above form, we obtain an asymmetric  DWCDD model, enabling a more realistic description of the protein conformation via the radius of gyration \(r\). As the initial condition for \(r\), as in the symmetric-potential case, we set $r(t)=r_{-}$ at $t=0$.

When employing a composite double-well potential, one must account for the fact that the escape time depends on which well the system escapes from. According to the Arrhenius law, the mean escape times from the wells at \(r_\pm\) are
\begin{align}
\mu_\pm \sim 2\,\alpha_\pm\,\delta_\pm\,\pi
\exp\!\left(\frac{V(r_{\mathrm M})-V(r_\pm)}{k_{\mathrm B}T}\right),
\label{eq:esc_time_DDwell}
\end{align}
where \(r_{\mathrm M}\) denotes the position of the barrier top.
Here, \(\alpha_{\pm}\) are prefactors determined by the curvatures of the potential in the vicinity of \(r=r_{\pm}\), and \(\delta_{+}\) (\(\delta_{-}\)) is a constant computed from the curvature at \(r=r_{\mathrm M}\) of the left (right) branch of the double-well potential.
Since Eq.~\eqref{eq:D_eff_lambda} holds generally for sojourn times with different means, by substituting the escape times given by Eq.~\eqref{eq:esc_time_DDwell}, the effective diffusion coefficient for an asymmetric potential can be computed as
\begin{widetext}
\begin{align}
    D_\mathrm{eff}=D_\mathrm{eq}-(D_+-D_-)\frac{\alpha_+\delta_+e^{\frac{U_{\rm L}}{k_{\rm B}T}}}{\alpha_+\delta_+e^{\frac{U_{\rm R}}{k_{\rm B}T}} + \alpha_-\delta_-e^{\frac{U_{\rm L}}{k_{\rm B}T}}}\left(\frac{1-e^{-t/\kappa(T)}}{t/\kappa(T)}\right),
    \label{eq:Deff_asym}\\
    \kappa(T)
    \equiv
    2\,\alpha_{+}\delta_{+}\,\alpha_{-}\delta_{-}\,\pi e^{\frac{U_{\rm L}+U_{\rm R}}{k_{\rm B}T}}
    \left(\alpha_+\delta_+e^{\frac{U_{\rm L}}{k_{\rm B}T}}+\alpha_-\delta_-e^{\frac{U_{\rm R}}{k_{\rm B}T}}\right)^{-1}.
    \label{eq:kappa_asym}
\end{align}
\end{widetext}
Here, $U_{\rm L}$ and $U_{\rm R}$ denote the heights of the left and right potential barriers, respectively, defined as $U_{\rm L} = V(r_{\mathrm M}) - V(r_+)$ and $U_{\rm R} = V(r_{\mathrm M}) - V(r_-)$.

Equation~\eqref{eq:Deff_asym} elucidates how \(D_{\mathrm{eff}}\) relaxes toward equilibrium as a function of temperature $T$ or observation time $t$ when the diffusivity switches between wells of a double-well potential with unequal depths \(U_L\) and \(U_R\). The parameter \(\kappa(T)\) sets the relaxation time of the diffusion process and, unlike the symmetric-potential case, it depends explicitly on the depths of the two wells.

\subsubsection{Effect of Potential Asymmetry}

\begin{figure*}
  \centering
  \includegraphics[width=\textwidth]{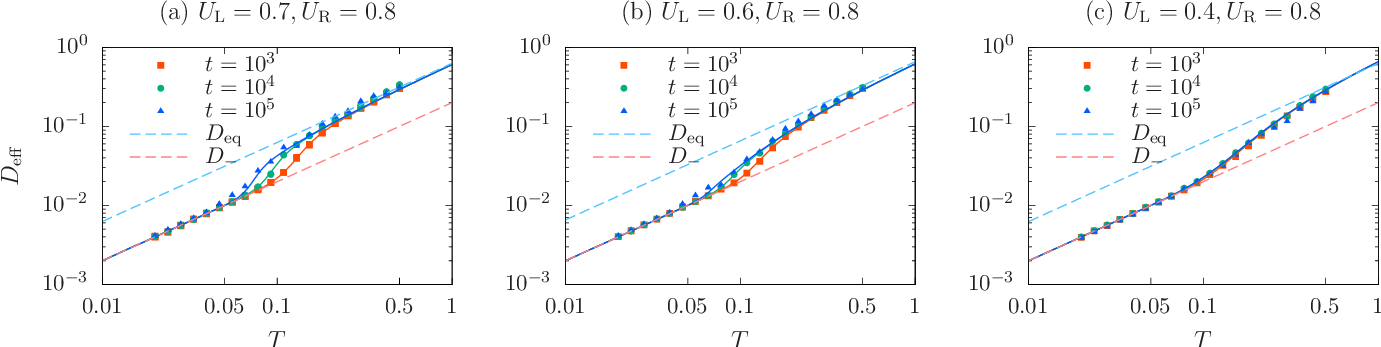}
  \caption{Temperature-dependent effective diffusion coefficient $D_\mathrm{eff}$ for different observation times in the asymmetric  DWCDD model. Each panel uses a different left-well depth \(U_\mathrm{L}\). Symbols indicate simulation results, dashed lines represent the initial and equilibrium diffusion coefficients, and solid lines show the theoretical prediction given by Eq.~\eqref{eq:Deff_asym}.}
  \label{fig:DT_asy}
\end{figure*}

Figure~\ref{fig:DT_asy} presents the temperature dependence of the effective diffusion coefficient \(D_{\mathrm{eff}}\) in the  DWCDD model with asymmetric well depths. In all cases, theoretical predictions from Eq.~\eqref{eq:Deff_asym} (solid lines) are in excellent agreement with the simulation results (symbols). When the asymmetry is weak, an observation-time-induced crossover is clearly observed and the transition temperature depends on the observation time (panel (a)). As the asymmetry increases, this time dependence progressively weakens, and the transition temperature becomes essentially independent of the observation time (panels (b) and (c)). This can be explained as follows. As the potential asymmetry increases ($U_\mathrm{L}\ll U_\mathrm{R}$), the equilibrium distribution of the particle radius \(r\) approaches a Gaussian peaked at \(r=r_{-}\). In the strongly asymmetric limit, the equilibrium distribution can be approximated by a delta function \(\delta(r-r_{-})\), which is nearly identical to the initial distribution. Consequently, the third condition for an observation-time-induced crossover discussed in the main text---“The system is initially out of equilibrium.”---is no longer satisfied, and the transition behavior ceases to depend on the observation time. Therefore, even in asymmetric two-state systems, an observation-time-induced crossover emerges, and the three conditions identified above should be regarded as necessary conditions for the occurrence of this phenomenon.

\subsection{Extraction of relaxation dynamics from trajectory data}

To illustrate how the present framework provides quantitative access to
underlying conformational dynamics, we show how relaxation properties can be inferred from finite-time measurements of effective diffusivity. Observation-time-induced crossovers arise only when the system is initialized in a non-equilibrium state; in the diffusivity model, this is implemented by fixing the initial radius in one of the basins, such as the low-diffusivity minimum.  Molecular-dynamics (MD) simulations of proteins, however, typically begin from an equilibrium ensemble \cite{Yamamoto2021}. When relaxation dynamics are to be probed, the initial state can nevertheless be controlled explicitly---for example, by selecting configurations from a particular basin (compact or expanded) or by briefly equilibrating the system at a chosen temperature. Thus, although observing the crossover requires specifying an initial basin, such initialization can be straightforwardly imposed in MD simulations or single-particle tracking protocols.

Importantly, the effective diffusivity contains direct information about the underlying conformational landscape. For a double-well-type landscape, the ratio of radii in the two basins is related to the diffusivity ratio through
\begin{align}
\frac{r_{-}}{r_{+}} = 2\frac{D_{\mathrm{eq}}}{D_{-}} - 1 ,
\end{align}
which is exact for a symmetric potential and remains a good approximation whenever the escape times from the two minima are of the same order. Thus, the magnitude of conformational change can be directly inferred from the measured diffusivity contrast.
MD simulations typically report radius-of-gyration changes of about $r_-/r_+ \simeq 1.2$ \cite{Yamamoto2021}; inserting this value gives $D_{\rm eq}/D_- \simeq 1.1$. Hence even a modest ($\sim$10\%) change in effective diffusivity signals a realistic conformational transition.

The temperature dependence of the crossover also encodes the energy barrier. A transition becomes apparent when the observation time $t_{\rm obs}$ matches the relaxation time $\kappa(T)$ governing interconversion between basins. For thermally activated dynamics,
\begin{equation}
t_{\rm obs} \sim \kappa(T_c) = \alpha\delta\pi\,\exp\!\left(\frac{\Delta U}{k_{\mathrm B}T_c}\right),
\end{equation}
so that
\begin{align}
    \log (t_{\mathrm{obs}}) \;\sim\; \frac{\Delta U}{k_{\mathrm B}}\,\frac{1}{T_c} \;+\; \log(\alpha\delta\pi).
\end{align}
Thus, by measuring the systematic shift of the apparent crossover temperature with observation time, one can directly estimate the conformational barrier $\Delta U$.

Together, these relations demonstrate that finite-time diffusivity data provide quantitative access to the characteristic radii and barrier heights governing conformational relaxation---even when only trajectory-level observables are available.

\subsection{Relation to the dynamical transition and experimental implications}

The protein DT observed in neutron-scattering experiments is generally associated
with a matching between the instrumental observation window and the relaxation
times of internal motions, often discussed in terms of
hydration-water hydrogen-bond rearrangements \cite{chen2006observation,schiro2015translational}.
In this kinetic picture, the apparent transition temperature reflects the point at
which characteristic internal relaxation processes become accessible within the
finite observation time.
Our results show that the same timescale-matching condition underlies the
observation-time-induced crossover produced by fluctuating diffusivity.

A key requirement for this crossover is that the system is initially biased away from the equilibrium diffusivity distribution. In DT experiments, samples are not prepared in a deliberately chosen non-equilibrium basin as in our protocol. Nevertheless, the relevant internal degrees of freedom often relax on timescales much longer than the picosecond-nanosecond window probed by neutron scattering. Thus, after a temperature change these modes cannot fully equilibrate within the measurement time and remain biased toward their pre-transition state, providing an effective non-equilibrium initial condition.

\begin{figure*}
  \centering
  \includegraphics[width=1.0\linewidth]{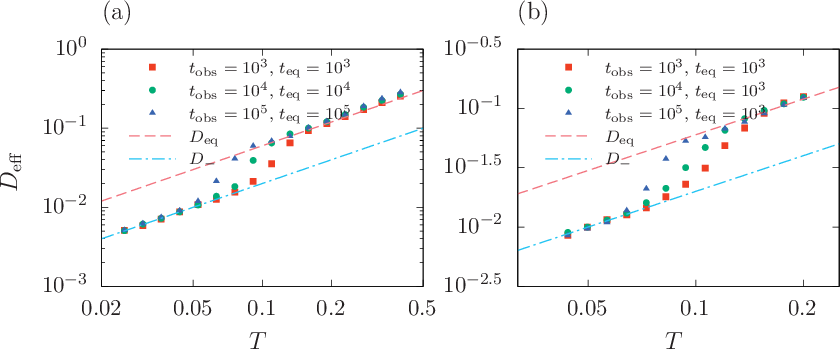}
  \caption{
(a) Temperature dependence of the effective diffusion coefficient in the DWCDD model when the system is equilibrated for a time equal to the observation time $t_{\rm obs}$. 
(b) Temperature dependence of the effective diffusion coefficient when the equilibration time is fixed to $t_{\mathrm{eq}} = 10^3$, independently of $t_{\rm obs}$. 
In both panels, the dashed lines indicate the initial effective diffusivity $D_-$ and the equilibrium effective diffusivity $D_{\mathrm{eq}}$.}
  \label{fig:DT_DWCDDModel_t_eq}
\end{figure*}

\subsubsection{ Effective non-equilibrium initial condition and finite equilibration time}

Importantly, this situation persists even when a finite equilibration time $t_{\rm eq}$ is allowed prior to measurement. As long as $t_{\rm eq}$ and the observation window are not sufficient to erase the initial bias, finite-time measurements probe a partially relaxed state and the crossover remains observable.
Therefore, DT measurements generically realize an effective non-equilibrium initial condition for the relevant internal modes, so the requirement for the observation-time-induced crossover is naturally met.

To explicitly demonstrate this robustness against finite equilibration,
we performed numerical simulations of the DWCDD model with controlled
equilibration protocols, as shown in Fig.~\ref{fig:DT_DWCDDModel_t_eq}. In these simulations, 
the particle radius is initially fixed at the low-diffusivity minimum $r = r_-$, followed by an equilibration stage of duration $t_{\mathrm{eq}}$ before the measurement of the effective diffusion coefficient.

Figure~\ref{fig:DT_DWCDDModel_t_eq}(a) shows the temperature dependence of the effective diffusivity when the equilibration time is chosen equal to the observation time, $t_{\mathrm{eq}} = t_{\mathrm{obs}}$. Despite this equilibration stage, a clear crossover remains visible, and the apparent crossover temperature continues to depend on the observation time. This demonstrates that equilibration on the same timescale as the measurement is insufficient to suppress the observation-time-induced crossover.

Figure~\ref{fig:DT_DWCDDModel_t_eq}(b) presents results obtained with a fixed equilibration time $t_{\mathrm{eq}} = 10^{3}$, independent of the observation time. Even in this case, the effective diffusivity exhibits a pronounced observation-time-dependent crossover. Together, these results show that the crossover is robust against finite equilibration and persists whenever slow relaxation prevents full equilibration prior to measurement.

\subsubsection{ Heating-cooling asymmetry}
This picture also implies a heating-cooling asymmetry. Upon heating from a low-mobility basin, relaxation is slow at low temperatures, so a non-equilibrium bias persists during the measurement and the observation-time-induced crossover appears. By contrast, at high temperature the diffusivity relaxes rapidly and its distribution quickly reaches the stationary form. Once equilibrated, subsequent cooling does not regenerate a meaningful non-equilibrium bias; rather, the system tracks the equilibrium distribution quasi-adiabatically at each temperature. Therefore the non-equilibrium initial condition required for the crossover is absent in the cooling protocol, and no crossover is expected. 
To directly test this scenario, we performed simulations using
a sequential temperature protocol mimicking heating and cooling experiments.

In our baseline protocol, at each temperature we imposed the initial condition $r = r_-$ and $x = 0$ independently for every realization. Here we instead modify the protocol so that the temperature is varied sequentially while observations are performed.
The simulation procedure is as follows. At the initial temperature $T = T_0$, we set the initial condition at $t = 0$ to be $r = r_-$ and $x = x_0$. After evolving the system at $T_0$, we change the temperature to the next value $T = T_1$ without reinitializing $r$ or $x$: the values of $r$ and $x$ at the end of the run at $T_0$ are directly used as the initial condition at $t = 0$ for the simulation at $T_1$. Repeating this procedure yields, for a sequentially measured set of $N$ temperatures $\{T_0, T_1, \dots, T_{N-1}\}$, the effective diffusion coefficient defined from the mean-square displacement as
\begin{equation}
    D_{\mathrm{eff}}(t) \equiv \frac{\bigl\langle \bigl(x(t) - x(0)\bigr)^2 \bigr\rangle}{2t}.
\end{equation}
By choosing the sequence $\{T_0, T_1, \dots, T_{N-1}\}$ in ascending or descending order, we can probe the effect of heating and cooling protocols, respectively.

\begin{figure}[t]
  \centering
  \includegraphics[width=\linewidth]{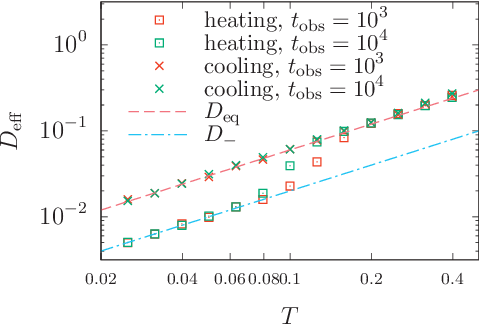}
  \caption{
Temperature dependence of the effective diffusion coefficient for the heating and cooling protocols. Squares denote the heating protocol and crosses denote the cooling protocol. The dashed lines indicate the initial effective diffusivity $D_-$ and the equilibrium effective diffusivity $D_{\mathrm{eq}}$. The initial temperature in the heating protocol is $T_0 \simeq 0.025$, while that in the cooling protocol is $T_0 \simeq 0.4$.
}
  \label{fig:heating_cooling}
\end{figure}

Figure~\ref{fig:heating_cooling} shows the temperature dependence of the effective diffusion coefficient obtained from simulations following the protocol described above. For the heating protocol, relaxation is slow in the low-temperature regime, so the influence of the non-equilibrium initial condition persists during the measurement and the effective diffusion coefficient exhibits a crossover as a function of temperature. Moreover, when the observation time is varied, the temperature at which this crossover occurs shifts, demonstrating the observation-time-induced crossover.

By contrast, for the cooling protocol, although we impose a non-equilibrium initial condition $r = r_-$ at the starting temperature $T_0$, the diffusivity relaxes rapidly in the high-temperature regime. Once the system has equilibrated at high temperature, subsequent cooling does not regenerate a significant non-equilibrium bias; instead, the system quasi-adiabatically follows the equilibrium distribution at each temperature, and the effective diffusion coefficient remains at its equilibrium value throughout the temperature range.

This scenario suggests an experimental asymmetry: DT should be suppressed, or even absent, under cooling protocols in which the system remains near equilibrium at each temperature.
More generally, heating and cooling need not follow symmetric relaxation pathways.
Recent studies have shown that relaxation dynamics can depend strongly on the
direction of temperature change, even in the absence of any thermodynamic phase
transition \cite{ibanez2024heating,Tejero2025}.

Beyond its connection to internal dynamical transitions, the present framework
also makes a concrete and experimentally accessible prediction for translational
transport.
Specifically, the observation-time-induced crossover should be directly observable
in the temperature dependence of the center-of-mass diffusion coefficient
extracted from finite-time trajectories.

While neutron-scattering experiments primarily probe internal atomic motions,
center-of-mass diffusion can be accessed in molecular-dynamics simulations and
single-particle tracking experiments.
Our theory predicts that the effective center-of-mass diffusivity measured over
a finite observation window exhibits a sharp temperature-dependent crossover,
whose apparent transition temperature systematically shifts with the observation
time and depends on the thermal protocol.

Such measurements provide a direct route to testing the proposed mechanism,
independently of microscopic details of internal motions.
In particular, by varying the observation time and the heating or cooling protocol,
one can extract the relaxation timescale governing the crossover, thereby linking
finite-time transport measurements to underlying slow conformational dynamics.

\section{Conclusion}
In summary, we have shown that fluctuating diffusivity in the Langevin framework generically leads to an observation-time-induced crossover, in which the effective diffusion coefficient exhibits a sharp, temperature-dependent transition whose crossover point shifts systematically with the observation time. By combining analytical treatments of a coarse-grained two-state model with numerical simulations of a  DWCDD model, we identified the minimal conditions required for the emergence of such transitions: (i) temporal fluctuations of diffusivity, (ii) temperature-dependent relaxation times, and (iii) non-equilibrium initial conditions. Our results identify nonequilibrium initial conditions as a necessary ingredient 
for the emergence of observation-time-induced crossover, 
thereby clarifying a key element in the dynamical interpretation of the DT.

Beyond protein dynamics, the present framework applies broadly to systems in which internal degrees of freedom modulate particle mobility while relaxing on temperature-dependent timescales. Examples include polymeric systems with fluctuating conformations, soft and glassy materials exhibiting slow structural relaxation, and active or driven systems where mobility is coupled to internal states. Because the mechanism does not rely on equilibrium assumptions and remains operative even in nonequilibrium steady states, it provides a unified viewpoint for interpreting finite-time transport measurements across a wide class of complex systems.

More generally, our results highlight the importance of observation protocols—such as the choice of observation time, initial preparation, and thermal history—in analyzing temperature-dependent transport data. Apparent transitions in effective diffusivity need not signal underlying thermodynamic phase changes, but can instead reflect kinetic crossovers governed by relaxation dynamics. We expect that this perspective will be useful in reinterpreting experimental and simulation data in soft matter, biophysics, and nonequilibrium statistical mechanics, and in extracting relaxation timescales and energy barriers from finite-time diffusion measurements.

\section*{Acknowledgements}
The authors thank Tomoshige Miyaguchi and Eiji Yamamoto for fruitful discussions. 
T.A. was supported by JSPS Grant-in-Aid for Scientific Research (No.~C 21K033920).

\appendix
\section{Observation-Time-Induced Crossover in non-equilibrium Steady States (Three-State Model)}
In this section, we extend the two-state model to a three-state model and show that a dynamical crossover arises even in cases where the instantaneous diffusion coefficient does not converge to its equilibrium value. The instantaneous diffusion coefficient takes three discrete values
\begin{align}
    D(t)\in\{D_1,D_2,D_3\},\quad D_3>D_2>D_1.    
\end{align}
Furthermore, the transition probabilities between states are specified by the transition matrix
\begin{align}
    P=\begin{pmatrix}
    0 & \tfrac{1}{3} & \tfrac{2}{3} \\
    \tfrac{2}{3} & 0 & \tfrac{1}{3} \\
    \tfrac{1}{3} & \tfrac{2}{3} & 0
    \end{pmatrix}
\end{align}
Here, \(P_{ij}\) denotes the probability of transitioning from state \(i\) to state \(j\).
The sojourn time in each state is modeled, as in the two-state model, by the exponential distribution
\begin{equation}
    \rho(\tau)=\frac{1}{\mu} e^{-\tau/\mu},\quad \tau\ge 0.
\end{equation}
Here, \(\mu\) denotes the mean residence time in each states and is assumed to follow an Arrhenius-type temperature dependence given by
\begin{align}
\mu(T)=\tau_0 \exp\!\left(\frac{\Delta U}{k_{\mathrm B}T}\right).
\end{align}
Here, \(\tau_0\) is a characteristic residence time that is independent of \(\Delta U\) and \(T\).
As the initial condition, we set the instantaneous diffusion coefficient to $D(0)=D_1$.

Under the non-equilibrium initial condition, the instantaneous diffusion coefficient can be expressed as
\begin{align}
\langle D(t)\rangle
&= D_1 + (D_2 - D_1)\,p_2(t) + (D_3 - D_1)\,p_3(t),\label{eq:D_t_three}
\end{align}
where \(p_2(t)\) and \(p_3(t)\) are the probabilities of finding the system in states 2 and 3 at time \(t\), respectively.
Using renewal theory, the probabilities are obtained as
\begin{align}
p_{2}(t)&=\frac{1}{3}
   - \frac{2}{3}\exp\!\left(-\frac{3}{2} \frac{t}{\mu}\right)
     \sin\!\left(\omega t+\frac{\pi}{6}\right),\label{eq:p2_t}\\
p_{3}(t)&=\frac{1}{3}
   + \frac{2}{3}\exp\!\left(-\frac{3}{2}\frac{t}{\mu}\right)
     \sin\!\left(\omega t-\frac{\pi}{6}\right)\label{eq:p3_t},
\end{align}
with \(\omega=(2\sqrt{3}\mu)^{-1}\).
Therefore, by substituting these results into Eq.~\eqref{eq:D_t_three} and integrating over time, the effective diffusion coefficient is obtained as
\begin{widetext}
\begin{align}
    D_{\mathrm{eff}}
    = D_{\mathrm{st}}
    + 2(D_2-D_1)\,
       \frac{\tfrac{1}{2}
       - B_-(t,\omega)\exp\!\left[-\tfrac{3t}{2\tau_0}e^{-\Delta U/(k_{\mathrm B}T)}\right]}
       {7t\,\tau_0^{-1}e^{-\Delta U/(k_{\mathrm B}T)}}
    + 2(D_3-D_1)\,
       \frac{1
       - B_+(t,\omega)\exp\!\left[-\tfrac{3t}{2\tau_0}e^{-\Delta U/(k_{\mathrm B}T)}\right]}
       {7t\,\tau_0^{-1}e^{-\Delta U/(k_{\mathrm B}T)}} .\label{eq:D_eff_three}
\end{align}
\end{widetext}
Here, the steady-state effective diffusion coefficient is
\begin{align}
    D_{\mathrm{st}}=\frac{D_1+D_2+D_3}{3},
\end{align}
and the oscillatory terms are
\begin{align}
B_{\pm}(t,\omega)
= \frac{3}{2}\sin\!\left(\omega t \pm \frac{\pi}{6}\right)
  + \frac{1}{2\sqrt{3}}\cos\!\left(\omega t \pm \frac{\pi}{6}\right).
\end{align}
Equation~\eqref{eq:D_eff_three} describes how a system whose diffusion coefficient switches among three states separated by a potential barrier of height \(\Delta U\) relaxes from a non-equilibrium initial condition to the steady state. This follows directly from the behavior of \(p_2(t)\) and \(p_3(t)\) in Eqs.~\eqref{eq:p2_t} and \eqref{eq:p3_t}, which oscillate while decaying exponentially; consequently, the instantaneous diffusion coefficient relaxes as the state-occupancy probabilities approach their stationary value \(1/3\) for each state.

\begin{figure}[t]
    \begin{center}
        \includegraphics[width=.8\linewidth]{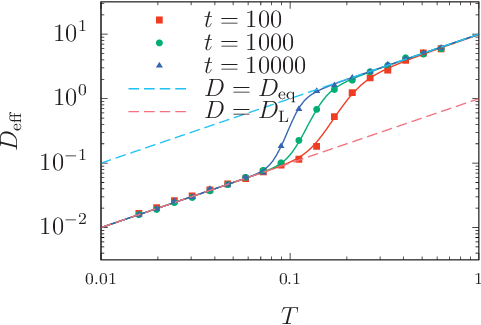}
        \caption{Temperature-dependent effective diffusion coefficient for different observation times in the three-state model. Symbols indicate simulation results, dashed lines represent the initial and equilibrium diffusion coefficients, and solid lines show the theoretical prediction given by Eq.~\eqref{eq:D_eff_three}. In the simulations, we set \(\tau_0=1\) and \(\Delta U=1\).}
        \label{fig:DT_threestate}
    \end{center}
\end{figure}

Figure~\ref{fig:DT_threestate} shows the temperature dependence of the effective diffusion coefficient in the three-state model. The results clearly demonstrate an observation-time-induced crossover. Even when the instantaneous diffusion coefficient has no equilibrium value, if the state distribution reaches stationarity at high temperatures or over sufficiently long observation times, the diffusion of particle positions becomes practically indistinguishable from that in equilibrium systems. Consequently, the effective diffusion coefficient exhibits a temperature crossover that depends on the observation time. In contrast, at low temperatures or over short observation times, biases from the initial condition persist and non-equilibrium behavior remains pronounced.

These results also demonstrate that the emergence of an observation-time-induced crossover requires not merely switching between two states, but the presence of two or more transition states in the instantaneous diffusion coefficient.

\section{Observation-Time-Induced Crossover in Systems with Non-Markovian Internal Dynamics}
In the main text, for analytical tractability, we assumed that the internal radius variable $r(t)$ follows a Markovian dynamics. In realistic complex systems, however, internal degrees of freedom often exhibit non-Markovian relaxation due to coarse graining, many-body interactions, or viscoelastic effects.

In this Appendix, we employ a generalized Langevin framework for $r(t)$ and demonstrate by numerical simulations that the observation-time-induced crossover remains robust even when the internal dynamics are non-Markovian.
Within this framework, the time evolution of the internal coordinate $r(t)$ is described by the generalized Langevin equation (GLE)
\begin{align}
    m\frac{d^{2}r}{dt^{2}}
    = -\int_{0}^{t}\Gamma(t-t')\,v(t')\,dt'
    -\frac{\partial V(r)}{\partial r}
    +\zeta(t),
    \label{eq:GLE_r}
\end{align}
where $v(t)\equiv \dot r(t)$, $\Gamma(t)$ is the memory kernel, and $\zeta(t)$ is a zero-mean Gaussian noise whose correlations are related to $\Gamma(t)$ via the fluctuation--dissipation theorem,
\begin{align}
\begin{split}
    \langle \zeta(t)\rangle &= 0,\\
\langle \zeta(t)\zeta(t')\rangle &= k_{\rm B}T\,\Gamma(|t-t'|).
\label{eq:FDT_GLE}    
\end{split}
\end{align}
Equation~\eqref{eq:GLE_r} is a non-Markovian generalization of Eq.~\eqref{eq:LE_r} in the main text, and it reduces to Eq.~\eqref{eq:LE_r} in the Markovian limit.
In the GLE, we assume an exponential memory kernel of the form
\begin{align}
\Gamma(t)=\gamma\lambda\,e^{-\lambda t}\qquad (t\ge 0),
\label{eq:exp_memory_kernel}
\end{align}
where $\gamma$ sets the overall friction strength and $\lambda^{-1}$ is the memory time.

For numerical simulations, it is convenient to rewrite the non-Markovian dynamics
\eqref{eq:GLE_r} as an equivalent Markovian system.
For the exponential kernel \eqref{eq:exp_memory_kernel}, we generate a colored random force
by an Ornstein--Uhlenbeck (OU) process
\begin{align}
\dot S(t)= -\lambda S(t) + \lambda \sqrt{2k_{\rm B}T\,\gamma}\,\xi_0(t),
\label{eq:OU_S}
\end{align}
where $\xi_0(t)$ is a Gaussian white noise with $\langle\xi_0(t)\rangle=0$ and
$\langle\xi_0(t)\xi_0(t')\rangle=\delta(t-t')$.
This OU process is chosen so that it satisfies the fluctuation--dissipation relation
\eqref{eq:FDT_GLE}; in particular,
\begin{align}
\langle S(t)S(t')\rangle
= k_{\rm B}T\,\gamma\lambda\,e^{-\lambda|t-t'|}
= k_{\rm B}T\,\Gamma(|t-t'|).
\label{eq:SS_corr}
\end{align}
Next, we introduce an auxiliary variable $u(t)$ by
\begin{align}
\sqrt{\lambda}\,u(t)
\equiv
-\int_{0}^{t}\gamma\lambda e^{-\lambda(t-t')}\,v(t')\,dt'
+S(t),
\label{eq:def_u}
\end{align}
which absorbs the memory integral.
Differentiating Eq.~\eqref{eq:def_u} yields
\begin{align}
\dot u(t)
= -\gamma\sqrt{\lambda}\,v(t) - \lambda u(t)
+ \sqrt{2\lambda k_{\rm B}T\,\gamma}\,\xi_0(t).
\label{eq:u_dyn}
\end{align}
As a result, the GLE \eqref{eq:GLE_r} with the exponential memory kernel
\eqref{eq:exp_memory_kernel} is equivalent to the following Markovian set of
stochastic differential equations
\begin{align}
\dot r(t) &= v(t), \label{eq:markov_r}\\
m\dot v(t) &= -\frac{\partial V(r)}{\partial r} + \sqrt{\lambda}\,u(t), \label{eq:markov_v}
\end{align}
together with Eq.~\eqref{eq:u_dyn}, which closes the dynamics.

In our simulation,
we set the friction strength and the memory parameter to $\gamma=5$ and $\lambda=0.01$.
Near the minimum $r=r_-$, we approximate the double-well potential as
$V(r)\simeq V(r_-)+\kappa_-(r-r_-)^2/2$ with $\kappa_-\equiv V''(r_-)$.
For the quartic potential \eqref{eq:DWpotential} with the parameters used in the main text,
we obtain $\kappa_-=V''(5)=1.6$.
In the Markovian limit, the overdamped intra-well relaxation time is then estimated as
\(\tau_{\rm well}\sim \gamma/\kappa_- \simeq 3.1\), 
whereas the memory time is
\(
\tau_{\rm mem}=\lambda^{-1}=100.
\)
Since $\tau_{\rm mem}\gg \tau_{\rm well}$, the friction retains a long memory over times far exceeding the
intrinsic relaxation of $r(t)$ within a basin. Since $\tau_{\rm mem}\gg \tau_{\rm well}$, memory effects are expected to be significant in the relaxation of $r(t)$.

\begin{figure}[t]
        \includegraphics[width=1.0\linewidth]{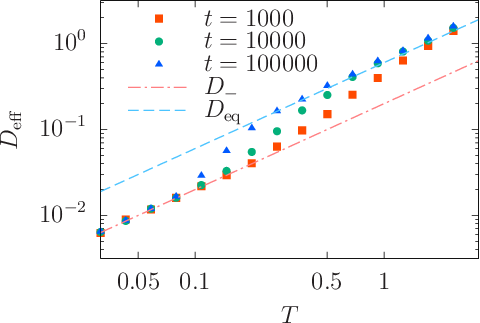}
        \caption{Temperature dependence of the effective diffusion coefficient $D_{\mathrm{eff}}(t)$ of the particle position $x(t)$ when the internal coordinate $r(t)$ follows the generalized Langevin dynamics \eqref{eq:GLE_r} with the exponential memory kernel \eqref{eq:exp_memory_kernel}. Symbols show simulation results for different observation times $t_{\mathrm{obs}}$, while the dashed lines indicate the initial value $D_-$ and the long-time (equilibrium) value $D_{\mathrm{eq}}$. A clear observation-time-induced crossover is observed: the apparent crossover temperature shifts systematically with $t_{\mathrm{obs}}$ even in the non-Markovian case ($\lambda=0.01$).}
        \label{fig:DT_GLE_exp}
\end{figure}

Figure~\ref{fig:DT_GLE_exp} shows the temperature dependence of the effective diffusion coefficient of the particle position $x(t)$ when the internal coordinate $r(t)$ follows the GLE dynamics.
As seen in the figure, $D_{\mathrm{eff}}$ exhibits a clear crossover as a function of temperature even for non-Markovian $r(t)$, and the apparent crossover temperature shifts systematically with the observation time.
These results indicate that the crossover is robust against non-Markovian internal relaxation: the Markovian assumption for $r(t)$ in the main text is made only for analytical convenience and is not essential for its emergence.

\section{Generality for various sojourn-time distributions}
In this section, we demonstrate that the observation-time-induced crossover arises for a broad class of sojourn-time statistics in the two-state model, including stretched-exponential (Weibull) distributions and power-law-tailed distributions.
In particular, as long as the characteristic timescale of the sojourn time depends on temperature, the observation-time-induced crossover generically emerges for a wide range of sojourn-time distributions.

The relaxation of the effective diffusion coefficient toward its equilibrium value $D_{\mathrm{eq}}$ is expected to occur when the observation time becomes comparable to the relaxation time of the diffusion state, which is the key timescale behind the observation-time-induced crossover.
For a generic two-state switching process, the relaxation time of the diffusion state is of the same order as the typical sojourn time.
Therefore, the deviation of $D_{\mathrm{eff}}(t)$ from $D_{-}$ sets in at times much shorter than the typical sojourn time of the underlying distribution.
Accordingly, by evaluating an approximate expression of $D_{\mathrm{eff}}(t)$ in the short-time regime, we can reproduce the initial rise of the diffusion behavior.
The ensemble average of the instantaneous effective diffusivity $\langle D_{\mathrm{eff}}(t)\rangle$ can be written in terms of the state probability $p_{+}(t)$ [Eq.~\eqref{eq:Dtwithp+}]. 
Within renewal theory~\cite{Cox, God2001, Miyaguchi2016, Akimoto2023}, the Laplace transform of $p_{+}(t)$ is given by
\begin{align}
  \hat{p}_{+}(s)
  =\frac{\hat{\rho}_{-}(s)}{s}\,
  \frac{1-\hat{\rho}_{+}(s)}{1-\hat{\rho}_{+}(s)\hat{\rho}_{-}(s)} .
  \label{eq:pplus}
\end{align}
Since the short-time limit $t\to0$ corresponds to $s\to\infty$, we may approximate $\hat{p}_{+}(s)\sim \hat{\rho}_{-}(s)/s$.
Accordingly, for small $t$ the state probability is estimated as
\begin{align}
  p_{+}(t)\sim \int_{0}^{t}\rho_{-}(t')\,dt',
  \label{eq:CDF}
\end{align}
namely, by the cumulative distribution function (CDF) of the $-$ state sojourn-time distribution.
In other words, in this regime the probability of having undergone at least one renewal up to time $t$ is dominated by the first sojourn in the initial $-$ state.
\subsection{Weibull sojourn times.}
We first consider Weibull-distributed sojourn times, whose probability density function is given by
\begin{align}
  \rho_{\pm}(\tau)
  = \frac{k}{\lambda_{\pm}}
  \left(\frac{\tau}{\lambda_{\pm}}\right)^{k-1}
  \exp\!\left[-\left(\frac{\tau}{\lambda_{\pm}}\right)^{k}\right], \quad \tau \geq 0,
  \label{eq:weibull_pdf}
\end{align}
where $k>0$ is the shape parameter and $\lambda_{\pm}>0$ is the scale parameter for the $\pm$ state.
We assume an Arrhenius-type temperature dependence for the scale parameter,
\begin{align}
  \lambda_{\pm}(T)
  = \tau_{\rm A}^{\pm}\exp\!\left(\frac{E_{0}^{\pm}}{k_{\mathrm{B}}T}\right),
  \label{eq:lambda_Tdep}
\end{align}
where $\tau_{\rm A}^{\pm}$ is the prefactor timescale and $E_{0}^{\pm}$ is the activation energy for the $\pm$ state.

Since the $-$-state CDF is $F_{-}(t)=1-\exp[-(t/\lambda_{-})^{k}]$, we have the short-time approximation
$p_{+}(t)\sim (t/\lambda_{-})^{k}$ as $t\to0$.
Using this approximation in Eq.~\eqref{eq:MSD}, we obtain
\begin{align}
  D_{\mathrm{eff}}(t)
  \sim D_{-}+\frac{D_{+}-D_{-}}{k+1}\left(\frac{t}{\lambda_{-}}\right)^{k}.
  \label{eq:Deff_weibull_short}
\end{align}
Equation~\eqref{eq:Deff_weibull_short} makes it explicit that the deviation from $D_{-}$ is governed by the dimensionless ratio $t/\lambda_{-}$.
For a fixed observation time $t$, increasing $T$ decreases $\lambda_{-}(T)$ and hence enhances the correction term $\propto (t/\lambda_{-})^{k}$, leading to an earlier departure of $D_{\mathrm{eff}}(t)$ from $D_{-}$.
Substituting the Arrhenius-type temperature dependence Eq.~\eqref{eq:lambda_Tdep} into Eq.~\eqref{eq:Deff_weibull_short}, we finally arrive at
\begin{align}
  D_{\mathrm{eff}}(t;T)
  \sim D_{-}+\frac{D_{+}-D_{-}}{k+1}
  \left(\frac{t}{\tau_{\rm A}^{-}\exp\!\left(\frac{E_{0}^{-}}{k_{\mathrm{B}}T}\right)}\right)^{k}.
  \label{eq:Deff_weibull_short_Tdep}
\end{align}
Equation~\eqref{eq:Deff_weibull_short_Tdep} describes how, for a fixed observation time $t$, the effective diffusion coefficient departs from the nonequilibrium value $D_{-}$ as the temperature is increased, reflecting the progressive equilibration of the diffusion state.

\begin{figure}[t]
  \centering
  \includegraphics[width=1.0\linewidth]{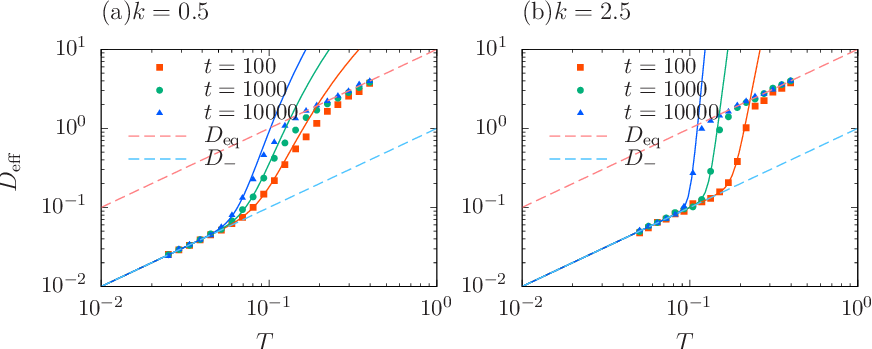}
  \caption{Temperature dependence of the effective diffusion coefficient $D_{\mathrm{eff}}(t)$ in the two-state model with Weibull-distributed sojourn times [Eq.~\eqref{eq:weibull_pdf}]. Symbols show simulation results, while the solid line represents the short-time prediction in Eq.~\eqref{eq:Deff_weibull_short_Tdep}. Panels (a) and (b) correspond to $k=0.5$ and $k=2.5$, respectively. In the simulations, we set $D_-=1$ and $D_+=19$, and chose the Arrhenius parameters as $\tau_{\rm A}^{\pm}=1$ and $E_{0}^{\pm}=1$.
}
  \label{fig:weibull_Deff_T}
\end{figure}

Figure~\ref{fig:weibull_Deff_T} shows the temperature dependence of the effective diffusion coefficient in the two-state model when the sojourn-time distribution is chosen as the Weibull form in Eq.~\eqref{eq:weibull_pdf}.
The simulation results (symbols) exhibit a clear crossover of $D_{\mathrm{eff}}$ as a function of temperature, and the initial rise is well captured by the prediction of Eq.~\eqref{eq:Deff_weibull_short_Tdep}.
The shape parameter $k$ controls the sharpness of the onset: for $k>1$ the correction term in Eq.~\eqref{eq:Deff_weibull_short_Tdep} grows more steeply with $t/\lambda_{-}(T)$, whereas for $0<k<1$ the departure from $D_{-}$ is more gradual.
Thus, while the detailed form of the rise depends on the sojourn-time statistics, the observation-time-induced crossover itself remains robust and is observed across different waiting-time distributions as long as the characteristic timescale varies with temperature.
\subsection{Power-law sojourn times (Lomax distribution).}
We next consider power-law-tailed sojourn times described by the Lomax (Pareto type-II) distribution, whose probability density function is given by
\begin{align}
  \rho_{\pm}(\tau)
  = \frac{\alpha}{\tau_{0}^{\pm}}
  \left(1+\frac{\tau}{\tau_{0}^{\pm}}\right)^{-(1+\alpha)}, \quad \tau \geq 0,
  \label{eq:lomax_pdf}
\end{align}
where $\alpha>0$ is the tail exponent and $\tau_{0}^{\pm}>0$ sets the characteristic timescale for the $\pm$ state.
We assume an Arrhenius-type temperature dependence for this timescale,
\begin{align}
  \tau_{0}^{\pm}(T)
  = \tau_{\rm A}^{\pm}\exp\!\left(\frac{E_{0}^{\pm}}{k_{\mathrm{B}}T}\right),
  \label{eq:tau0_Tdep}
\end{align}
where $\tau_{\rm A}^{\pm}$ is the prefactor timescale and $E_{0}^{\pm}$ is the activation energy for the $\pm$ state.

The cumulative distribution function (CDF) of the $-$-state sojourn time is
\begin{align}
  F_{-}(t)
  = 1-\left(1+\frac{t}{\tau_{0}^{-}}\right)^{-\alpha}.
  \label{eq:lomax_cdf_minus}
\end{align}
Using the short-time approximation $p_{+}(t)\simeq F_{-}(t)$ and substituting it into Eq.~\eqref{eq:MSD}, we obtain
\begin{widetext}
\begin{align}
  D_{\mathrm{eff}}(t)\sim
  \begin{cases}
    D_{-}+(D_{+}-D_{-})\left[
      1-\dfrac{\tau_{0}^{-}}{t}\,
      \dfrac{\left(1+t/\tau_{0}^{-}\right)^{1-\alpha}-1}{1-\alpha}
    \right],
    & (\alpha\neq 1),\\
    D_{-}+(D_{+}-D_{-})\left[
      1-\dfrac{\tau_{0}^{-}}{t}\ln\!\left(1+\dfrac{t}{\tau_{0}^{-}}\right)
    \right],
    & (\alpha=1).
  \end{cases}
  \label{eq:Deff_lomax_cases}
\end{align}
\end{widetext}
Substituting Eq.~\eqref{eq:tau0_Tdep} into Eq.~\eqref{eq:Deff_lomax_cases} yields explicit temperature dependences through $\tau_{0}^{-}(T)$.
As in the Weibull case, the deviation from $D_{-}$ is governed by the dimensionless ratio $t/\tau_{0}^{-}(T)$, while the detailed onset profile depends on the sojourn-time statistics.

\begin{figure}[t]
  \centering
  \includegraphics[width=1.0\linewidth]{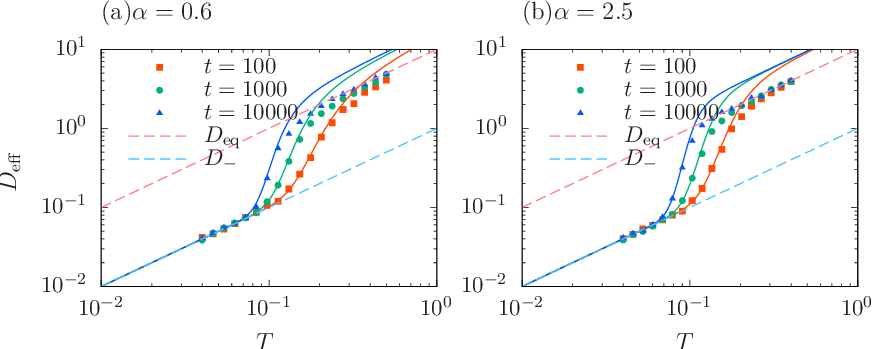}
  \caption{Temperature dependence of the effective diffusion coefficient $D_{\mathrm{eff}}(t)$ in the two-state model with Lomax-distributed sojourn times [Eq.~\eqref{eq:lomax_pdf}]. Symbols show simulation results, while the solid line represents the short-time prediction in Eq.~\eqref{eq:Deff_lomax_cases}. Panels (a) and (b) correspond to $\alpha=0.6$ and $\alpha=2.5$, respectively. In the simulations, we set $D_-=1$ and $D_+=19$, and chose the Arrhenius parameters as $\tau_{\rm A}^{\pm}=1$ and $E_{0}^{\pm}=1$.
}
  \label{fig:lomax_Deff_T}
\end{figure}

Figure~\ref{fig:lomax_Deff_T} shows the temperature dependence of the effective diffusion coefficient in the two-state model when the sojourn-time distribution is chosen as the Lomax form in Eq.~\eqref{eq:lomax_pdf}.
The simulation results (symbols) exhibit a clear crossover of $D_{\mathrm{eff}}$ as a function of temperature, and the initial rise is well captured by the prediction of Eq.~\eqref{eq:Deff_lomax_cases}. 

The Lomax distribution is characterized by a heavy power-law tail. As a consequence, its second moment diverges for $\alpha<2$, and even the mean sojourn time diverges for $\alpha<1$.
Despite these anomalous properties, the simulation results for $\alpha=0.6$ in Fig.~\ref{fig:lomax_Deff_T} still show a clear observation-time-induced crossover in $D_{\mathrm{eff}}$.
This is because the emergence of the crossover is controlled primarily by the short-time buildup of the escape probability from the initial slow state, which depends on the dimensionless ratio $t/\tau_0^{-}(T)$ rather than on the existence of higher-order moments of the sojourn-time distribution.
Accordingly, even in the regime of divergent mean sojourn times, a temperature-dependent characteristic scale $\tau_0^{-}(T)$ is sufficient to produce an observation-time-induced crossover.

These results highlight that the observation-time-induced crossover does not rely on Markovian switching or finite moments, but instead on the temperature-dependent timescale governing the early-time departure from the initial condition.

%

\end{document}